\documentclass[aps, pre, reprint, twocolumn, floatfix]{revtex4-2}
\usepackage{amsmath, amssymb}
\usepackage{graphicx}
\usepackage{xcolor}
\usepackage{booktabs}
\usepackage{lipsum}
\usepackage{balance}
\usepackage{hyperref}

\begin{document}

\title{Engineering Polarization:\\ How Contradictory Stimulation Systematically Undermines Political Moderation}

\author{Renato Vieira dos Santos}
\affiliation{Universidade Federal de Lavras (UFLA), Instituto de Ciência, Tecnologia e Inovação (ICTIN), R. Antonio Carlos Pinheiro Alcântara, n° 855 - Jardim Mediterranée, São Sebastião do Paraíso, MG, 37950000, Brazil}

\date{\today}

\begin{abstract}
Political moderation, a key attractor in democratic systems, proves highly fragile under realistic information conditions. We develop a stochastic model of opinion dynamics to analyze how noise and differential susceptibility reshape the political spectrum. Extending Marvel et al.'s deterministic framework, we incorporate stochastic media influence $\zeta(t)$ and neuropolitically-grounded sensitivity differences ($\sigma_y > \sigma_x$). Analysis reveals the moderate population---stable in deterministic models---undergoes catastrophic collapse under stochastic forcing. This occurs through an effective deradicalization asymmetry ($u_{B}^{\text{eff}} = u + \sigma_y^2/2 > u_{A}^{\text{eff}}$) that drives conservatives to extinction, eliminating cross-cutting interactions that sustain moderates. The system exhibits a phase transition from multi-stable coexistence to liberal dominance, demonstrating how information flow architecture---independent of content---systematically dismantles the political center. Our findings reveal moderation as an emergent property highly vulnerable to stochastic perturbations in complex social systems.\\

\noindent\textbf{Keywords: }{Stochastic polarization dynamics, Political opinion formation, Moderation collapse, Socio-physical systems, Complex social dynamics}\\

\noindent\textbf{DOI:} \url{https://doi.org/10.1016/j.chaos.2025.117554}
\end{abstract}



\maketitle

\section{Introduction}
\label{s1}

The stability of democratic societies is an emergent property of a complex social system, intimately linked to the vitality of its political center. In the framework of sociophysics, moderate actors can be seen as a critical attractor state that integrates ideas from across the political spectrum, upholding the tradition of constructive conflict essential for preventing systemic polarization \cite{1,2,3,4,5,6,8,9}. Yet, a pervasive trend of political polarization threatens to erode this centrist foundation in many modern democracies \cite{3}, presenting a fundamental problem in the dynamics of social systems. Understanding the mechanisms that foster or undermine this stable, moderate state is therefore a question of profound importance for both societal resilience and the physics of complex systems.

Recent comprehensive reviews in social physics have further established the mathematical foundations for modeling complex social phenomena, including opinion formation and polarization dynamics \cite{59}. These developments highlight the growing recognition of physics-inspired approaches to understanding social systems. Concurrently, empirical studies have demonstrated how success-driven opinion formation can generate and amplify social tensions through reinforcement mechanisms \cite{60}, providing crucial context for our investigation into the structural vulnerabilities of political moderation.

In a seminal contribution to this field, Marvel et al. \cite{10,13,15,17,18} employed a simple yet powerful mathematical model from statistical physics to explore strategies for encouraging ideological moderation. Their model considered interactions between population classes as a dynamical system. From seven distinct strategies, only one proved successful: \textit{nonsocial deradicalization}. In this deterministic framework, a stable, mixed population including a substantial fraction of moderates can persist, representing a theoretical beacon of hope for mitigating polarization.

However, this successful strategy was identified under a critical and arguably simplistic assumption: a consistent, unambiguous pro-moderation signal. This idealized assumption stands in stark contrast to the stochastic reality of contemporary information ecosystems. The modern public sphere is characterized by high-volume, high-velocity information flows that are often fragmented, inconsistent, and contradictory \cite{47,48}---creating an environment of pervasive \textit{informational noise}. This noise is not processed uniformly, and a growing body of evidence from neuropolitics reveals fundamental differences in how liberals and conservatives process information, suggesting \textit{differential susceptibility} to chaotic signals \cite{12,14,16}.

In this paper, we investigate the \textit{fragility of this deradicalization attractor} under a more realistic, stochastic information environment. We extend the deterministic model into the stochastic regime by incorporating two key features: (i) \textit{Stochastic Media Influence}, modeled as a noisy drive $\zeta(t),$ and (ii) \textit{Differential Susceptibility}, parameterized from neuropolitical findings ($\sigma_y>\sigma_x$). Our analysis reveals that the previously stable population of moderates undergoes a \textit{catastrophic collapse} under the stochastic dynamics of contradictory stimulation, leading to outcomes of polarization or extinction. This result provides a formal demonstration of how the architecture of modern information flow---irrespective of content---can systematically engineer the collapse of the political center, a phenomenon that can be understood as a \textit{noise-induced phase transition} in a complex social system.

\section{Model Formulation and Stochastic Framework}

We formulate the opinion dynamics as a stochastic dynamical system, extending the deterministic framework of Marvel et al. \cite{10} into a regime with parametric noise and differential susceptibility. The model presented in this section is a conceptual tool designed to illustrate the core argument. While simple, it offers a formal framework to explore the implications of our hypotheses.

Let $A,$ $B,$ $AB,$ $A_c$ symbols denote the liberals, conservatives, moderates and committed liberals (militants), respectively. The following hypotheses will be considered (interpret symbols such as $A + B \to AB$ as ``speaker'' + ``listener'' $\to$ ``something''. $A+B\to AB$ has the interpretation: when a liberal ($A$) hears a conservative ($B$), there is a tendency of the liberal become more moderate ($AB$)):
\begin{enumerate}
    \item When a conservative (liberal) hears a liberal (conservative), there is a tendency of the conservative (liberal) becomes more moderate: $B+A\to AB,$ $(A+B\to AB).$ 
    \item When a moderate hears a conservative (liberal), there is a tendency of the moderate becomes more conservative (liberal): $AB+B\to B,$ $(AB+A\to A).$
    \item When a conservative hears a committed liberal, there is a tendency of the conservative becomes more moderate $B+A_c\to AB.$
    \item When a moderate hears a committed liberal, there is a tendency of the moderate becomes more liberal $AB+A_c\to A.$
    \item The total population is fixed, so that $n_A+n_B+n_{AB}+p=1,$ where $n_A,$ $n_B,$ $n_{AB}$ and $p$ denote the expected fractions of individuals corresponding to the liberals $A,$ conservatives $B,$ moderates $AB$ and committed liberals $p,$ respectively.
    \item The fraction $p$ of committed individuals is fixed and the initial conditions are $(n_A(0),n_B(0))=(0,1-p).$
\end{enumerate}

Table~\ref{tab:interactions} summarizes schematically some of the hypotheses:

\begin{table}[htbp]
\centering
\caption{Interaction rules in the opinion dynamics model}
\label{tab:interactions}
\begin{tabular}{ccccc}
\toprule
Speaker & + & Listener & $\to$ & Post-interaction \\
\midrule
$A$ & $+$ & $B$ & $\to$ & $AB$ \\
$A_c$ & $+$ & $B$ & $\to$ & $AB$ \\
$A_c$ & $+$ & $AB$ & $\to$ & $A$ \\
$AB$ & $+$ & $B$ & $\to$ & $B$ \\
$B$ & $+$ & $A$ & $\to$ & $AB$ \\
$AB$ & $+$ & $A$ & $\to$ & $A$ \\
\bottomrule
\end{tabular}
\end{table}

Note that committed liberals $A_c$ cannot be influenced. Based on these interaction rules and the \textit{law of mass action} \cite{7,20,21,22,24}, we can derive a system of equations describing the rate of change for each population:
\begin{equation}
\begin{split}
\dot{n}_A & = (p+n_A)n_{AB}-n_An_B, \\ 
\dot{n}_B & = n_Bn_{AB}-(p+n_A)n_B        
\end{split}
\label{eq1}
\end{equation}
with $n_{AB}=1-p-n_A-n_B$ and $\dot{X}\equiv dX/dt.$ These equations represent the backbone of the models analyzed in \cite{10}. Among the 7 variations proposed by Marvel et al., only one demonstrated the property that the population of moderates does not collapse or becomes subdued. This variant incorporates media effects and can be described mathematically as
\begin{equation}
\begin{split}
\dot{n}_A & = (p+n_A)n_{AB}-n_An_B-un_A, \\ 
\dot{n}_B & = n_Bn_{AB}-(p+n_A)n_B-un_B        
\end{split}
\label{eq2}
\end{equation}
Here, the parameter $u>0$ represents the rate at which the conservatives and liberals abandon their radical positions (or engages in it, $u<0$) in response to a \emph{nonsocial} stimulus. We interpret this nonsocial stimuli as being generated by \textit{mass media} mechanisms in general. This model was named \emph{nonsocial deradicalization.}

\begin{figure}[htbp]
\centering
\includegraphics[width=0.95\linewidth]{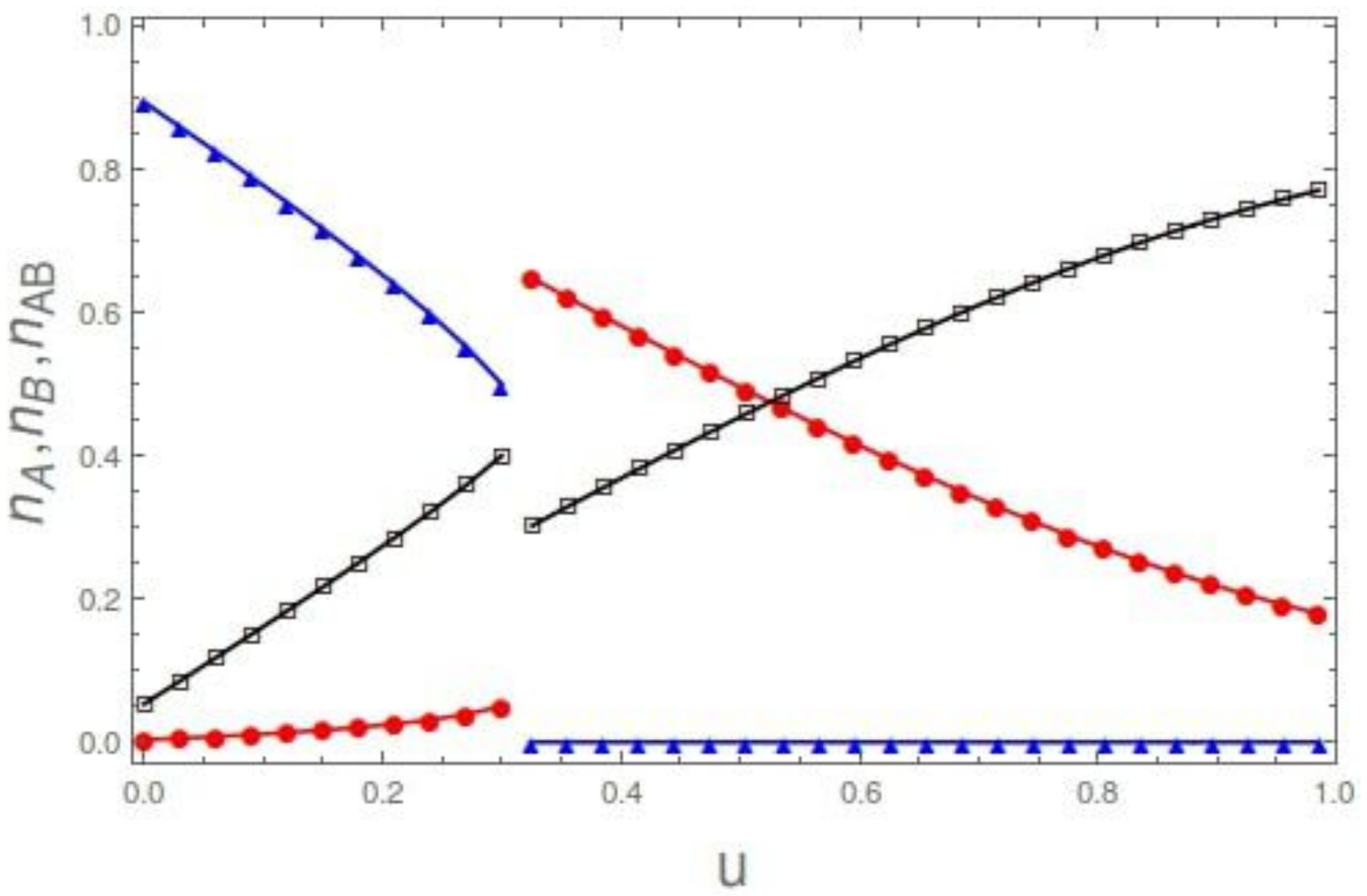}
\caption{Deterministic baseline: Population equilibria under nonsocial deradicalization. Shows how moderate populations (black squares) persist only within a specific range of media influence $u,$ while conservatives (blue triangles) are driven to extinction beyond a critical threshold $u_{cr},$ destabilizing the political ecosystem. Parameters: $n_{A_c}(0)=p,$ $n_{B}(0)=1-p$ with $p=5/100,$ $n_{AB}(0)=0$ and $n_{A}(0)=0,$ $u_{cr}=\frac{11}{10}-\sqrt{\frac{3}{5}}\approx0.325$.}
\label{fig1}
\end{figure}

The main outcome associated with this model is shown in Fig.~\ref{fig1} where we see how the equilibrium populations vary with the intensity of the mass media influence $u.$ Fig.~\ref{fig1} reveals a bifurcation in the system's equilibrium states as a function of the control parameter $u.$

There is a critical value of $u\equiv u_{cr}$ above which the population of conservatives is extinguished. For $0\leq u\leq u_{cr}$ the population of liberals increases slowly and after a subtle increment in $u=u_{cr},$ the population of liberals abruptly increases. The population of moderates grows with $0\leq u\leq u_{cr}$ and $u>u_{cr}$ with a little abrupt fall in $u=u_{cr}.$ The conclusion reached in \cite{10} is that the only way (out of 7 tested possibilities) to encourage the population of moderate is through pro-moderation mass media effort.\footnote{Note that despite the stimulus to the population of moderates, there is the extinction of the conservatives for $u>u_c.$} We will see that if we consider some results coming from the new field of \emph{Neuropolitics} \cite{11,19}, things can not be that simple.

\begin{figure*}[htbp]
\centering
\includegraphics[width=0.8\textwidth]{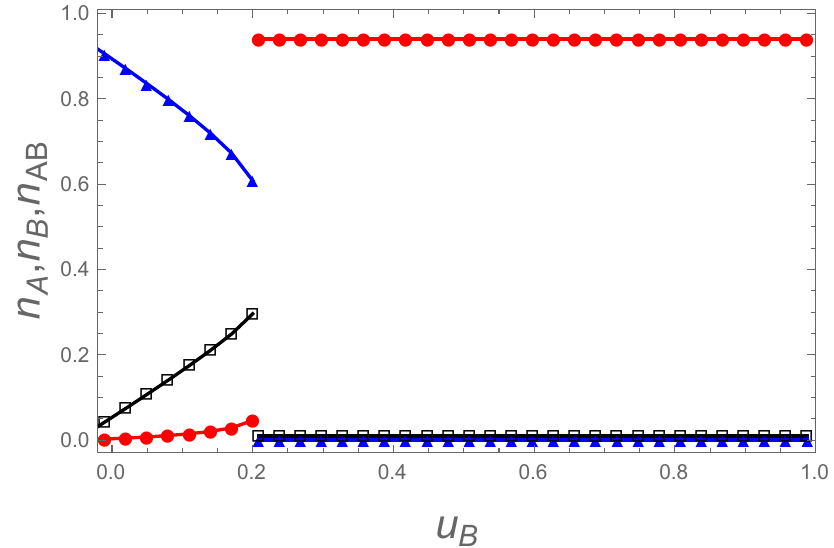}
\caption{Differential susceptibility in deterministic framework: Equilibrium populations as a function of conservative sensitivity $u_B$ for varying liberal sensitivities $u_A$. Clockwise from top-left: $u_A=\frac{1}{100}, \frac{1}{10}, \frac{1}{2}, 1$. The key insight reveals that lower liberal susceptibility to media influences (smaller $u_A$) systematically reduces equilibrium moderate populations (black squares), establishing the foundational asymmetry exploited in our stochastic extension. This demonstrates how inherent differences in media responsiveness can predispose the system to center collapse under noisy conditions.}
\label{sf}
\end{figure*}

Before discussing the recent discoveries of Neuropolitics, it is important to briefly mention the results for the equilibrium behavior of non-social deradicalization model considering different sensitivities to media stimuli. Let the model described by
\begin{equation}
\begin{split}
\dot{n}_A & = (p+n_A)n_{AB}-n_An_B-u_An_A, \\
\dot{n}_B & = n_Bn_{AB}-(p+n_A)n_B-u_Bn_B     
\end{split}
\label{eq3a}
\end{equation}
which may be written as\footnote{For analytical clarity in the subsequent stochastic analysis, we will adopt the variable mapping $n_A \rightarrow x, n_B \rightarrow y.$}
\begin{equation}
\begin{split}
\dot{x} & =  \left(1-2p-u_A\right)x-2xy - x^2 - p y + p - p^2, \\
\dot{y} & =  \left(1-2p-u_B\right) y-2 yx- y^2
\end{split}
\label{eq3b}        
\end{equation}
with $u_A$ and $u_B$ representing the different sensitivities to media stimuli of liberals and conservatives, respectively, and $n_A,$ $n_B,$ $n_{AB}$ are $x,$ $y$ and $z=1-p-x-y,$ respectively. The sequence of figures in Fig.~\ref{sf} illustrates what happens to the equilibrium populations if we vary $u_B$ for some specific values of $u_A.$

The relevant conclusion for this article is that \emph{low susceptibility of liberals to the media influences implies lower equilibrium values for the population of moderates.} This conclusion will be \emph{very important} later when we discuss the effects of \emph{contradictory stimulation}.

\section{Neuropolitical Foundations for Differential Susceptibility}

To ground the parameterization of differential susceptibility ($\sigma_y > \sigma_x$) in our stochastic extension, we require a plausible mechanistic hypothesis for how different groups process informational noise. We posit that established findings in neuropolitics provide a robust foundation for expecting systematic differences in susceptibility to contradictory stimulation. This section reviews the key evidence suggesting that conservatives and liberals exhibit differential sensitivity to the uncertainty and threat potentially embedded in a volatile information environment, justifying the core model assumption where $\sigma_y > \sigma_x$.

\subsection*{Neural Correlates of Uncertainty Tolerance}

A central dimension of our thesis involves tolerance for ambiguity and conflict. The anterior cingulate cortex (ACC), a brain region critically involved in conflict monitoring, error detection, and regulating responses to uncertainty, appears to be a key neuroanatomical site for relevant individual differences.

Kanai et al. \cite{16} provided a direct link between brain structure and political orientation by demonstrating that larger ACC grey matter volume was associated with greater liberalism. The authors hypothesize that this structural difference may be linked to a ``higher capacity to tolerate uncertainty and conflicts,'' a cognitive style more receptive to novel and complex information, which aligns with liberal ideologies. Conversely, a relatively smaller ACC volume in conservatives might indicate a lower threshold for perceiving and experiencing cognitive conflict, making inconsistent or ambiguous signals more psychologically taxing.

This finding is not isolated. Amodio et al. \cite{12} found that liberals showed significantly greater event-related ACC neural activity in response to cues signaling a need to change a habitual response---a task requiring conflict monitoring and cognitive flexibility. Conservatives, in contrast, displayed more stable neural activity in these situations. This suggests a neurocognitive basis for differing responses to information that challenges existing patterns or creates cognitive dissonance.

\subsection*{Physiological Sensitivity to Threat and Aversive Stimuli}

Beyond processing uncertainty, responses to directly threatening or aversive stimuli also show politically patterned differences. The amygdala, particularly the right amygdala, is a key structure in the neural circuitry for processing fear and emotional salience.

Multiple studies have linked conservatism to a heightened physiological responsiveness to threats. Oxley et al. \cite{14} found that individuals with measurably higher physiological reactions to sudden noises and threatening visual images were significantly more likely to support protective policies such as defense spending, capital punishment, and patriotism. Schreiber et al. \cite{15} further confirmed this link using neuroimaging, showing that during a risky decision-making task, Republican participants exhibited heightened amygdala activity compared to Democrats.

This sensitivity extends beyond direct physical threat to include moral disgust. Smith et al. \cite{17} demonstrated that individuals with marked involuntary physiological responses to disgusting images (e.g., a man eating worms) were more likely to self-identify as conservative and to adopt socially protective stances, such as opposing gay marriage. This suggests that a general sensitivity to aversive, norm-violating stimuli is a trait more commonly associated with conservatism.

\subsection*{Synthesis: From Neurophysiology to Model Parameterization}

These convergent lines of evidence paint a coherent picture: a neurocognitive profile associated with conservatism is characterized by lower tolerance for uncertainty and cognitive conflict (linked to ACC structure/function) and higher sensitivity to threatening and aversive stimuli (linked to amygdala reactivity).

In the context of our model, the parameter $\sigma$ represents the intensity of perceived ``contradictory stimulation'' from the media environment. We argue that this noise is not emotionally or cognitively neutral. A stochastic stream of conflicting narratives inherently contains high levels of uncertainty (What is true?) and potential symbolic threat (Which narrative, if believed, puts me or my values at risk?).

Therefore, it is a parsimonious and empirically grounded hypothesis that individuals with the neurocognitive traits more common among conservatives would experience a given level of informational noise ($\sigma$) as more salient, stressful, and impactful. This justifies our core modeling assumption of differential susceptibility, formalized as $\sigma_y > \sigma_x$. It is crucial to emphasize that this is not a value judgment on either orientation but a recognition of differential fit between inherent cognitive styles and the structure of the modern information environment. The subsequent section explores the dramatic consequences of this seemingly minor adjustment to the model of deradicalization.

\section{Results: The Stochastic Collapse of Moderation}

Having established the deterministic baseline and the neuropolitical rationale for differential susceptibility, we now present the outcomes of our stochastic extension. Our central question is: How robust is the ``nonsocial deradicalization'' strategy when the pro-moderation media signal is contaminated by noise, and this noise is perceived asymmetrically?

The original deterministic system for the ``nonsocial deradicalization'' strategy is given by:
\begin{equation}
\begin{split}
\dot{n}_A &= (p + n_A)n_{AB} - n_A n_B - u n_A, \\
\dot{n}_B &= n_B n_{AB} - (p + n_A)n_B - u n_B,        
\end{split}
\label{eq:deterministic}
\end{equation}
where $n_{AB}=1-p-n_A-n_B,$ and $u>0$ is the constant deradicalization rate. As shown in Fig.~\ref{fig1}, this system supports a stable equilibrium with a non-zero population of moderates $n_{AB}$ for a range of $u$ values.

We introduce stochasticity and differential susceptibility by transforming the media influence parameter $u$ into a stochastic variable, unique to each group:
\begin{equation}
u \rightarrow u + \zeta_i(t), \quad \text{for } i \in \{A, B\}.
\label{eq:stochastic_transform}
\end{equation}

Here, $\zeta_i(t)$ is a Gaussian white noise process with zero mean and variance $\sigma_i^2,$ such that $\langle\zeta_i(t)\rangle=0$ and $\langle\zeta_i(t')\zeta_i(t'')\rangle=\sigma_i^2\delta(t''-t')$. The key model extension is the hypothesis, grounded in Section 3, that $\sigma_y>\sigma_x$ indicating conservatives' higher sensitivity to the contradictory stimulation represented by the noise.

The resulting system of Stratonovich Stochastic Differential Equations (SDEs) is:
\begin{equation}
\begin{split}
dn_A &= \left[(p + n_A)n_{AB} - n_A n_B - u n_A\right]dt - n_A dB_t^A, \\
dn_B &= \left[n_B n_{AB} - (p + n_A)n_B - u n_B\right]dt - n_B dB_t^B,
\end{split}
\label{eq:SDEs}
\end{equation}
where $dB_t^{i}$ are the Wiener increments for groups $A$ and $B,$ correlated with their respective noise intensities. We adopt the Stratonovich interpretation of stochastic calculus, which is physically natural for systems with smooth, real-world noise \cite{40,41,42,45}.

\subsection{Analytical Insight via Moment Equations}

To gain analytical insight, we derive the approximate moment equations using a derivative matching closure method \cite{46,59}, which neglects temporal correlations ($\langle xy\rangle \approx \langle x\rangle\langle y\rangle$; see Appendix). We obtain:
\begin{equation}
\begin{split}
\dot{x} &= \left(1 - 2p - u - \frac{\sigma_x^2}{2}\right)x - 2xy - x^2 - p y + p - p^2, \\
\dot{y} &= \left(1 - 2p - u - \frac{\sigma_y^2}{2}\right)y - 2yx - y^2.
\end{split}
\label{eq:moment_equations}
\end{equation}

A critical observation is that these equations are mathematically equivalent to the original deterministic model under the transformation $u_A^{\text{eff}}=u+\sigma_x^{2}/2$ and $u_B^{\text{eff}}=u+\sigma_y^{2}/2$. This reveals that the effect of stochastic influence is isomorphic to an effective increase in the deradicalization rate in the deterministic model. Since $\sigma_y>\sigma_x,$ this implies $u_B^{\text{eff}}>u_A^{\text{eff}}.$ The original model showed that a sufficiently high deradicalization rate $u$ leads to the extinction of conservatives and an eventual drop in moderates. Our analysis thus predicts that differential susceptibility will selectively and asymmetrically impact the population dynamics, pushing the system towards a new equilibrium devoid of conservatives and with a fragile moderate population.

\subsection{Numerical Simulation of the Stochastic Dynamics}

The analytical approximation, while insightful, neglects temporal correlations and extinction events. To capture the full stochastic dynamics, we performed numerical simulations of the SDEs (Eqs.~\ref{eq:SDEs}) using the Euler-Maruyama method \cite{40,57}.

First, in the symmetric noise case ($\sigma_x=\sigma_y$), when both groups are equally susceptible to noise, the system's behavior moderates. While the average populations resemble the deterministic case, the stochasticity introduces fluctuations around these equilibria. The moderate population persists, though its size shows increased variance.

Second, in the differential susceptibility case ($\sigma_y>\sigma_x$), introducing our core hypothesis produces a dramatic shift. The conservative population $n_B,$ subjected to a higher effective deradicalization pressure $u_B^{\text{eff}},$ is driven towards extinction. Concomitantly, the moderate population $n_{AB},$ which previously thrived, collapses.

\begin{figure}[htbp]
\centering
\includegraphics[width=0.95\linewidth]{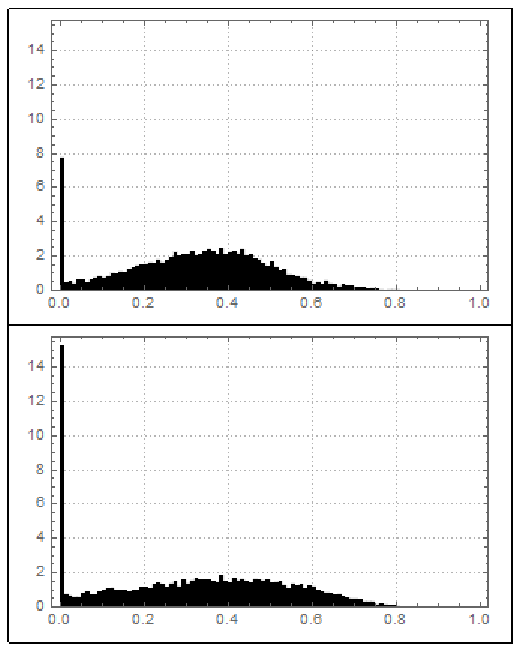}
\caption{Stochastic collapse of political moderation: Quasi-stationary probability density functions for the moderate population under differential noise susceptibility. (Top) For low conservative sensitivity ($\sigma_x=0.6$, $\sigma_y=0.25$), the distribution maintains a peak away from zero, indicating viable moderate coexistence. (Bottom) For high conservative sensitivity ($\sigma_x=0.6$, $\sigma_y=0.75$), the distribution undergoes catastrophic collapse to a peak at zero, demonstrating moderator extinction. This visualizes the central finding of our study---a noise-induced phase transition where the political center becomes unsustainable under asymmetric contradictory stimulation.}
\label{fig:Moderados_Sigma}
\end{figure}

Fig.~\ref{fig:Moderados_Sigma} illustrates this starkly. For low $\sigma_y$ (e.g., $\sigma_y=0.25$), the PDF for moderates is centered away from zero, indicating their survival. However, for high $\sigma_y$ (e.g., $\sigma_y=0.75$), the PDF for moderates becomes sharply peaked at zero. This signifies that in the long run, the most probable outcome is the extinction of the moderate population. The system evolves towards a state dominated almost entirely by liberals.

\begin{figure}[htbp]
\centering
\includegraphics[width=0.95\linewidth]{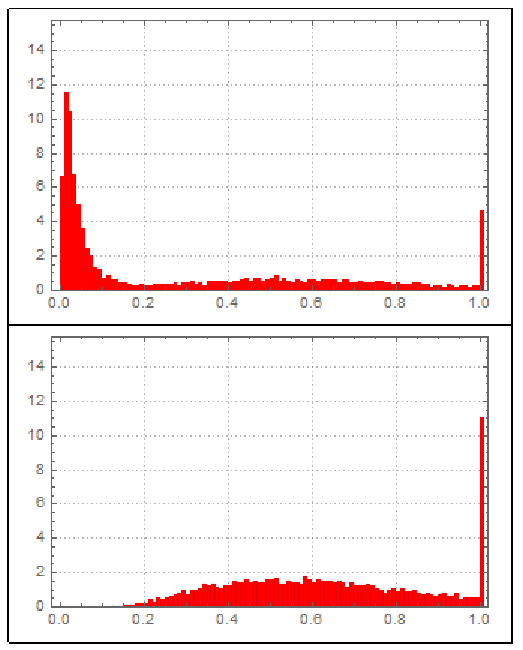}
\caption{Emergent liberal dominance following center collapse: Quasi-stationary probability density functions for the liberal population. (Top) $\sigma_x=0.6$, $\sigma_y=0.25$; (Bottom) $\sigma_x=0.6$, $\sigma_y=0.75$. As conservative sensitivity to contradictory stimulation increases, the liberal distribution shifts rightward and sharpens, indicating demographic dominance. Crucially, this liberal expansion occurs not through conversion of moderates but through the collapse of the cross-cutting interaction network that sustained the political center, revealing the ecosystem-level consequences of asymmetric noise susceptibility.}
\label{fig:Vermelhos_Sigma}
\end{figure}

Fig.~\ref{fig:Vermelhos_Sigma}, showing the PDF for the liberal population $(n_A),$ confirms the mechanism. As conservative sensitivity $(\sigma_y)$ increases, the liberal distribution shifts rightward and sharpens, indicating that liberals come to dominate the population with higher probability and less uncertainty.

\subsection{Interpretation of the Collapse}

In summary, the results demonstrate that the ``nonsocial deradicalization'' strategy is not robust to the introduction of realistic, asymmetrically perceived noise. The mechanism is twofold: First, the higher susceptibility of conservatives $(\sigma_y>\sigma_x)$ acts as a selective pressure, effectively applying a stronger deradicalization force to them and leading to their demographic extinction. Second, the moderates $(AB),$ who rely on cross-cutting interactions with both liberals $(A)$ and conservatives $(B)$ for their existence (e.g., via $A+B\rightarrow AB$), cannot be sustained in a political ecosystem where one of their foundational interaction partners has been eliminated. The moderate population, deprived of its ``ecological niche,'' subsequently collapses.

In conclusion, the only strategy proven to foster moderation in the deterministic model fails catastrophically under a stochastic regime that incorporates empirically grounded differences in neurocognitive processing. The moderate population, far from being stable, is shown to be a fragile phenomenon, highly susceptible to the structural noise of the modern information environment.

\section{Synthesis and Discussion: Resilience and Phase Transitions in Political Ecosystems}

Our stochastic analysis reveals that political moderation, while stable in deterministic frameworks, exhibits critical fragility under stochastic forcing---collapsing through a noise-induced phase transition. This section interprets this demographic shift through classical sociological frameworks (anomie), examines its weaponization via digital platforms, and explores implications for building democratic resilience in high-noise information environments.

\subsection{The Silent Extinction: Noise-Induced Phase Transition}

The analytical insight from moment equations predicted a fundamental shift: differential susceptibility ($\sigma_y > \sigma_x$) selectively targets conservatives with higher effective deradicalization, leading to their demographic extinction and consequent collapse of the moderate ecosystem.

The numerical simulations---specifically the quasi-stationary Probability Density Functions (PDFs) in Figs.~\ref{fig:Moderados_Sigma} and \ref{fig:Vermelhos_Sigma}---provide stark confirmation of this transition. Fig.~\ref{fig:Moderados_Sigma} shows the PDF for moderates ($n_{AB}$). At low conservative sensitivity ($\sigma_y = 0.25$), the distribution peaks away from zero, indicating metastable coexistence. At high sensitivity ($\sigma_y = 0.75$), the distribution undergoes a catastrophic collapse, peaking sharply at zero---a clear signature of a noise-induced phase transition where moderation becomes unsustainable. Fig.~\ref{fig:Vermelhos_Sigma}, showing liberal PDFs ($n_A$), confirms the mechanism. As $\sigma_y$ increases, the liberal distribution shifts rightward, indicating demographic dominance. Crucially, moderates don't convert to liberals; rather, the cross-cutting interactions that sustain them ($A + B \rightarrow AB$) become impossible when one interaction partner is eliminated.

This represents not ideological victory but \emph{ecosystem collapse}---the political center vanishes as collateral damage in an asymmetrical environmental stressor.

\subsection{From Model Dynamics to Sociological Reality: Complex Systems Perspective}

This mathematical collapse finds resonance in classical sociology. The model formalizes Durkheim's anomie \cite{34,35}---a state of normlessness emerging when integrative mechanisms fail. Here, the ``collective conscience'' is the cross-cutting discourse that moderates represent. Contradictory stimulation ($\sigma$), by making this discourse cognitively unsustainable, dissolves social cohesion.

The result is not new consensus but vacuum---a system dominated by a single attractor ($A$), with the former center vanished into political disengagement. Our model demonstrates how information architecture can engineer anomie, not by arguing against the center, but by making its maintenance prohibitively costly.

\subsection{The Hegelian Stress Test: When Dialectics Breaks Down}

This process represents a pathological instantiation of Hegelian dialectics. In healthy form, Thesis-Antithesis-Synthesis enables epistemic progress. When weaponized into continuous, high-frequency contradictory stimulation, the intended ``Synthesis'' becomes psychological breakdown.

Our model shows this manifests as center collapse. The moderate synthesis cannot hold under perpetual conflict. The resulting ``synthesis'' is the vacuum in our PDFs---a distribution where the middle disappears, leaving only a dominant pole and disengaged former centrists. This aligns with mass psychology insights \cite{21,25,26,27,28,29} that confusion and contradiction dissolve critical thought.

\subsection{The Engagement-Alignment Paradox: Platform Dynamics}

Our theoretical findings resonate powerfully with empirical digital landscapes. The model identifies high-$\sigma$ contradictory stimulation as the moderation collapse driver. Modern social platforms systematically maximize this parameter through their core mechanics:

First, algorithmic $\sigma$-maximization: Engagement-optimizing algorithms preferentially amplify content generating strong reactions---outrage, fear, tribal affirmation \cite{50,23,32,33}. Coherent, nuanced arguments (low-$\sigma$) are less ``engaging'' than conflicting, sensational claims (high-$\sigma$). Second, temporal and contextual collapse: Social media fragments complex discourse into simultaneous, decontextualized pieces---humanitarian appeals alongside genocidal rants, science alongside conspiracy. This represents the ``informational grille illusion'' where contradictory stimuli apply simultaneously, maximizing cognitive load \cite{36,37,58}. Third, personalized susceptibility amplification: Neurocognitive differences ($\sigma_y > \sigma_x$) become profiling and targeting variables. Micro-targeting can expose threat-sensitive users to more danger-related content, effectively personalizing and amplifying their $\sigma$ exposure \cite{38,39,49}.

Thus, social platforms function as $\sigma$-maximization engines. They operationalize dialectics not as philosophical process but as continuous psychological stress test. The predicted collapse serves engagement metrics while corroding democratic discourse.

\subsection{Implications: From Content Control to System Resilience}

These findings compel a paradigm shift: the primary threat is not extremist content but the chaotic information structure itself. Content-focused interventions resemble saving individual fish in an acidifying ocean---addressing symptoms while missing systemic pathology.

The model suggests instead fostering \emph{structural and cognitive resilience}---developing systems and citizens capable of processing noise without collapsing. This implies: First, epistemic stability: Education should prioritize critical thinking, logic, and media literacy as stable cognitive frameworks for navigating contradiction. Second, coherent deliberation spaces: The antidote to weaponized contradiction is more coherent speech---institutions and spaces dedicated to sustained civil dialogue create low-$\sigma$ havens where synthesis becomes possible. Third, center valorization: Cultural and political efforts should recognize the moderate, synthesizing mindset as the essential, fragile bedrock of pluralistic society.

\section{Conclusion: Complex Systems and Democratic Resilience}

This study demonstrates that political moderation constitutes an emergent property of stable systems, exhibiting critical fragility under stochastic contradictory stimulation. Extending a canonical model reveals how the only known solution for fostering moderates can be subverted by the very information environment it relies upon.

The social media ecosystem, in its current form, acts as a potent engine for maximizing the destructive parameter $\sigma$. However, a potential misinterpretation---advocating centralized control or censorship---would constitute a profound misdiagnosis. The weaponization of contradiction represents a pathology of disordered discourse, not free discourse itself.

The solution cannot lie in empowering central authorities to impose order---a cure destroying patient autonomy to save it. Such approaches replace chaotic digital noise with monolithic, state-sanctioned signals, replicating the totalitarian designs our work seeks to illuminate.

The model's dynamics point toward resilience principles found in complex adaptive systems: distributed cultural mechanisms and local interaction rules can generate system-level resistance to informational noise \cite{7,55}, while epistemic network analyses demonstrate that heterogeneous judgment and local knowledge processing enhance collective resilience to misinformation cascades \cite{47,50}. These distributed, bottom-up approaches to maintaining system stability stand in stark contrast to centralized information control, which often fails to capture the necessary complexity for genuine resilience \cite{56}.

This suggests a path both more challenging and more aligned with complex systems principles: First, intellectual patrimony: Resilience emerges not from blocking bad information but from furnishing minds with robust frameworks. Shared cultural and intellectual patrimony---philosophy, history, literature---provides stable reference points for navigating contradiction. Second, civil association: Local, voluntary communities act as essential low-$\sigma$ environments. These ``molecular structures'' of civil society \cite{52,53} train habits of reasoned dialogue and trust eroded by high-velocity digital discourse. Third, epistemic agency: The ultimate answer to engineered chaos is cultivating what Mill termed ``the sovereignty of the individual'' \cite{54}---educational and cultural ethos prizing epistemic agency, critical thinking, and independent judgment over algorithmic or state delegation.

In conclusion, our model serves not as control blueprint but as warning and renewal call. It reveals the center's fragility and how digital architectures actively test its limits. The democratic response cannot be simplistically turning down volume, but complexly learning to discern signal within noise---cultivating, through education, culture, and civil society, the intellectual and moral fortitude enabling free people to remain free amidst the storm.

\appendix
\section{Moment Equations Derivation}
\label{app1}

To obtain the equations for the moments, we use the equation
\begin{equation}
\frac{d\langle F(\textbf{x}(t),t)\rangle}{dt}=\left\langle \frac{\partial F}{\partial t}+\frac{\partial F}{\partial\textbf{x}}\cdot\textbf{a}+\frac{1}{2}Tr\left[\textbf{b}^T\cdot\frac{\partial^2F}{\partial\textbf{x}^2}\cdot\textbf{b}\right]\right\rangle,
\label{ap_eq1}
\end{equation}
where $\partial^2F/\partial\textbf{x}^2$ is the \emph{matrix} of second derivatives \cite{55}. For our problem, $\textbf{x}=\bigl[\begin{smallmatrix} x \\ y \end{smallmatrix} \bigr],$ $\textbf{a}=\bigl[\begin{smallmatrix} (p+x) (-p-x-y+1)-u x-x y \\ -y (p+x)+y (-p-x-y+1)-u y \end{smallmatrix} \bigr]$ and $\textbf{b}=\bigl[\begin{smallmatrix} \sigma_x x & 0 \\ 0 & \sigma_y y \end{smallmatrix} \bigr].$ If $F(\textbf{x})=\ln{x}$ and $F(\textbf{x})=\ln{y},$ the equations for $\langle\dot{x}\rangle$ and $\langle\dot{y}\rangle$ will be, respectively
\begin{equation}
\begin{split}
\langle\dot{x}\rangle & = \left(1-2p-u-\frac{\sigma_x^2}{2}\right)\langle x\rangle - 2\langle xy\rangle - \langle x^2\rangle - p\langle y\rangle \\&+ p-p^2 \\
\langle\dot{y}\rangle & = \left(1-2p-u-\frac{\sigma_y^2}{2}\right)\langle y\rangle-2\langle yx\rangle-\langle y^2\rangle
\end{split}
\label{ap_eq2}
\end{equation}
If $F(\textbf{x})=x^2,$ $F(\textbf{x})=y^2$ and $F(\textbf{x})=\ln{(xy)},$ we have
\begin{equation}
\begin{split}
\langle\dot{x^2}\rangle & = 2p(1-p)\langle x\rangle - 2p\langle xy\rangle   \\
& + (2-4p-2u+\sigma_x^2)\langle x^2\rangle - 2\langle x^3\rangle - 4\langle x^2y\rangle \\
\langle\dot{y^2}\rangle & = (2-4p-2u+\sigma_y^2)\langle y^2\rangle - 2\langle y^3\rangle - 4\langle xy^2\rangle \\
\langle\dot{xy}\rangle & = p(1-p)\langle y\rangle - 3\langle x^2y\rangle - 3\langle xy^2\rangle - p\langle y^2\rangle  \\
& + \left(2-4p-2u-\frac{\sigma_x^2}{2}-\frac{\sigma_y^2}{2}\right)\langle xy\rangle
\end{split}
\label{ap_eq3}
\end{equation}
To close the hierarchy of equations, we use the closure method called derivative matching \cite{56} and put $\langle x^3\rangle=\frac{\langle x^2\rangle^3}{\langle x\rangle^3},$ $\langle y^3\rangle=\frac{\langle y^2\rangle^3}{\langle y\rangle^3},$ $\langle x^2y\rangle=\frac{\langle x^2\rangle\langle xy\rangle^2}{\langle x\rangle^2\langle y\rangle}$ and $\langle xy^2\rangle=\frac{\langle y^2\rangle\langle yx\rangle^2}{\langle y\rangle^2\langle x\rangle}$ in the equations above:
\begin{equation}
\begin{split}
\langle\dot{x^2}\rangle & = 2p(1-p)\langle x\rangle - 2p\langle xy\rangle   \\
& + (2-4p-2u+\sigma_x^2)\langle x^2\rangle - 2\frac{\langle x^2\rangle^3}{\langle x\rangle^3} - 4\frac{\langle x^2\rangle\langle xy\rangle^2}{\langle x\rangle^2\langle y\rangle} \\
\langle\dot{y^2}\rangle & = (2-4p-2u+\sigma_y^2)\langle y^2\rangle - 2\frac{\langle y^2\rangle^3}{\langle y\rangle^3} - 4\frac{\langle y^2\rangle\langle yx\rangle^2}{\langle y\rangle^2\langle x\rangle} \\
\langle\dot{xy}\rangle & = p(1-p)\langle y\rangle - 3\frac{\langle x^2\rangle\langle xy\rangle^2}{\langle x\rangle^2\langle y\rangle} - 3\frac{\langle y^2\rangle\langle yx\rangle^2}{\langle y\rangle^2\langle x\rangle} - p\langle y^2\rangle  \\
& + \left(2-4p-2u-\frac{\sigma_x^2}{2}-\frac{\sigma_y^2}{2}\right)\langle xy\rangle
\end{split}
\label{ap_eq4}
\end{equation}
The system of equations (A2) and (A4) is closed and can be used in further analysis.

\balance

\end{document}